\documentclass[%
 reprint,
 amsmath,amssymb,
 aps,
]{revtex4-2}
\usepackage{graphicx}
\usepackage{dcolumn}
\usepackage{bm}

\begin{document}

\preprint{APS/123-QED}

\title{Engineering Si-Qubit MOSFETs: A Phase-Field Modeling Approach Integrating Quantum-Electrostatics at Cryogenic Temperatures}

\author{Nilesh Pandey$^1$}
\email{pandey@utexas.edu}
\author{Dipanjan Basu$^2$}

\author{Yogesh Singh Chauhan$^3$}
\author{Leonard F. Register$^1$}
\author{Sanjay K. Banerjee$^1$}
 \email{banerjee@ece.utexas.edu}
\affiliation{%
 $^1$Microelectronics Research Center, University of Texas at Austin, Austin, Texas\\
 $^2$Synopsys Inc., Hillsboro, OR, USA\\
 $^3$Indian Institute of Technology - Kanpur,  Kanpur, 208016, India 
}%

\begin{abstract}
This study employs advanced phase-field modeling to investigate Si-based qubit MOSFETs, integrating electrostatics and quantum mechanical effects. We adopt a comprehensive modeling approach, utilizing full-wave treatment of the Schrödinger equation solutions, coupled with the Poisson equation at cryogenic temperatures. Our analysis explores the influence of interface traps on quantum dot (QD) barrier heights,  affecting coupling due to tunneling. A wider trap distribution leads to the decoupling of quantum dots. Furthermore, the oscillations in the transmission \& reflection coefficients increase as the plunger/barrier gate length is increased, reducing the coupling between the QDs. 
By optimizing plunger and barrier gate dimensions, spacer configurations, and gap oxide lengths, we enhance control over quantum well depths and minimize unwanted wave function leakage. The modeling algorithm is also validated against the experimental data and can accurately capture the oscillations in the $I_d-V_{gs}$ caused by the Coulomb blockade at cryogenic temperature. 

\end{abstract}

\maketitle

\section{\label{sec:level1}Introduction}
Silicon-based qubit devices present exceptional scalability, targeting the future needs of millions of qubits for practical quantum computing [1], [2], [3], [4]. The electronics part of Si-based qubit systems has been discussed by Zwanenburg et. al. [1]. Recent advancements
in Si and SiGe quantum devices demonstrate exceptionally high low-temperature mobility [5], a six-bit quantum
processor [6], scalable silicon quantum processor design
using flip-flop qubit [7], greater than 0.1 ms relaxation
times [8], [9], silicon quantum dots with sufficient thermal
robustness to perform universal gate is demonstrated in
[10], quantum-state foundation as a tomography to characterize entanglement [11], two-bit CNOT Qubit logic
gate [12], FinFET based Si-qubit [13], 22-nm FD SOI
based single-electron injection qubit device [14], and high
fidelity silicon-based qubit [15], [16], [17].

There have been significant advancements in the qubit
simulation and modeling domain. Dynamical mean-field
theory for qubit [18], variational hybrid algorithm by using both classical and quantum physics [19], utilizing
molecular wave function to study qubit [20], quantum
monte-carlo simulations [21], first principle based simulations to analyze the nano-scale physics of qubit [22], [23],
[24] quantum simulation using fermionic swap network
[25], quantum imaginary time technique to determine
eigen energies [26], The physics of point defect and its impact on qubit is explored via experimental or simulations
[27], [28], [29], quantum phase transition [30], quantum
embed algorithm to simulate quantum computer [31], 2-
D material impact on qubit spin and electrostatics [32],
[33], [34].

From the large circuit perspective. Recently, a 3-D integration of a Si-qubit circuit was demonstrated in [35],
[36], [37]. A large-scale circuit such as the electronic controller was reported in [38], On-chip switch capacitor integration in [39], and a 28-nm node pulse modulator in
[40]. The recent good review articles discuss the advancement in qubit-based electronics from devices to circuits
[41], [42].
A recent study by Mohiyaddin et al. developed a multiphysics modeling algorithm to study Si-qubit devices [43].
However, the authors did not consider a full wave treatment, and the WKB approximation was used to calculate the tunnel coupling. The overlap of wave functions
among qubits is crucial for achieving multi-state storage,
superposition, and thoroughly understanding the electrostatics [1], [4]. Therefore, a detailed microscopic understanding of electron wave function interactions, tunneling, and exchange coupling is vital for efficient quantum
computer development. Additionally, Mohiyaddin et al.
work did not consider the impact of trap distribution and
plunger/barrier gate on the qubit device.
\begin{figure}[!t]
		\centering 
		\includegraphics[width=0.5\textwidth]{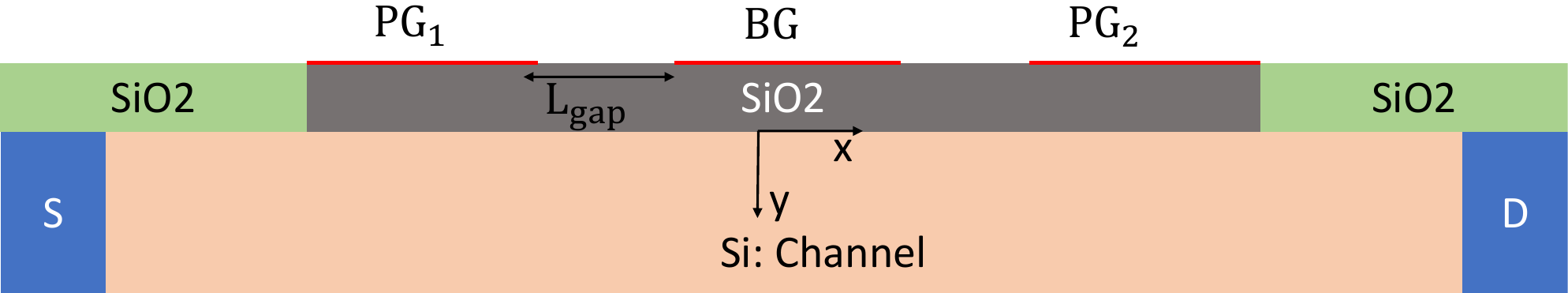}
		\caption{Schematic of a Si-based qubit MOSFET with two quantum dots (QD), highlighting the plunger gate (PG), barrier gate (BG), and gap oxide (L$_{gap}$) lengths. The default parameters are L$_{PG}$ = L$_{PG}$ = 40 nm, L$_{gap}$ = 10 nm, Oxide thickness (t$_{ox}$) = 3 nm, silicon film thickness (t$_{si}$) = 15 nm, S/D spacer length (SiO$_2$) = 50 nm, V$_{PG}=$ 1 V, V$_{BG}=$ 0.5 V, mid-gap work-function, and T = 10 K. 
		}\label{fig:schm}
	\end{figure}
 \begin{figure*}[!t]
		\centering 
		\includegraphics[width=0.8\textwidth]{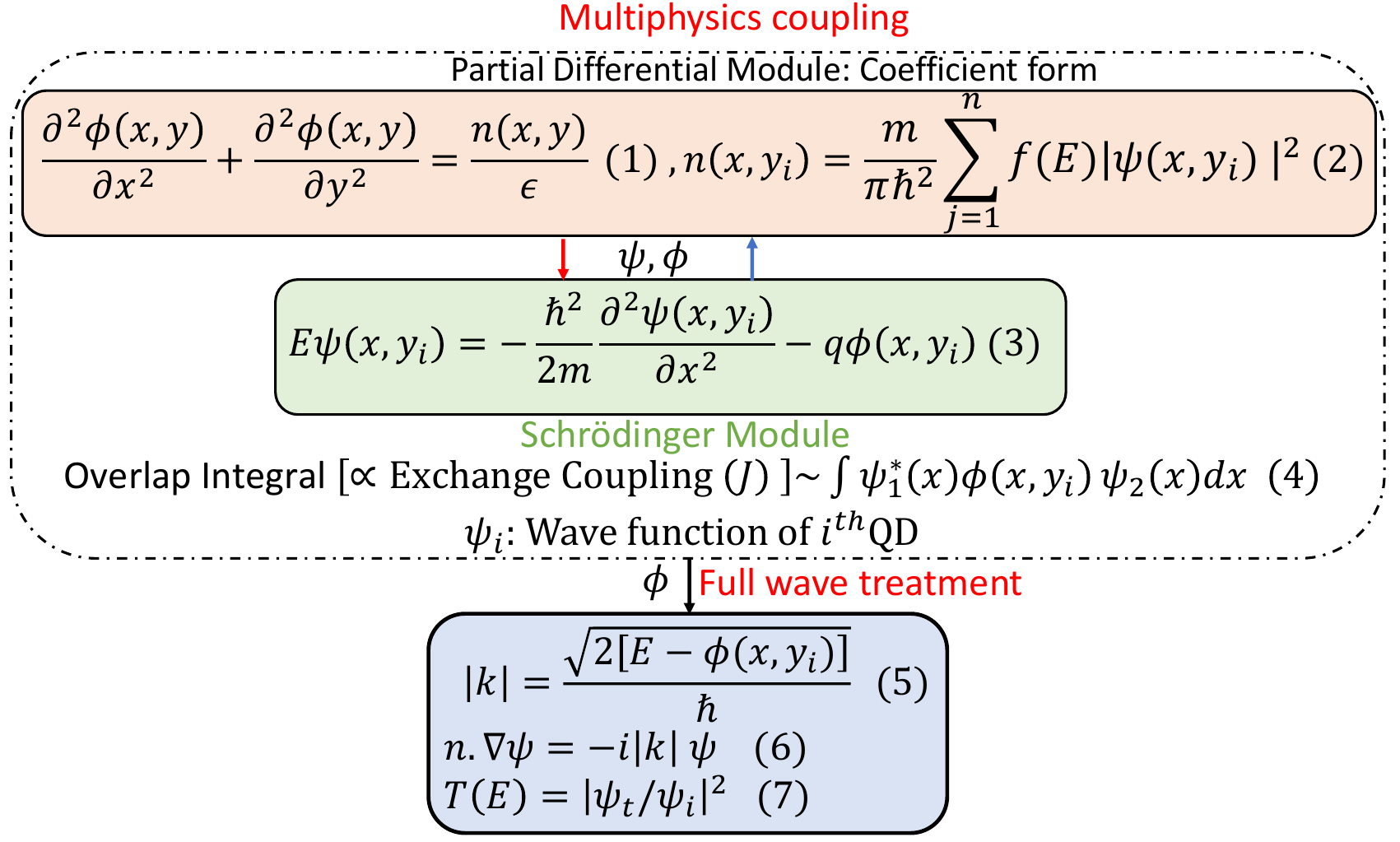}
		\caption{
  The phase-field modeling approach. It integrates electrostatics through the 2-D Poisson's equation, computes charge density using the full Fermi-Dirac distribution, and employs a 1-D Schrödinger equation for quantum mechanical effects, crucial for simulating Si-based qubit MOSFETs at cryogenic temperatures. Refer to the methodology section for details about the modeling and convergence at cryogenic temperature. Here, $y_i=y_s$ is used to analyze the electrostatic at SiO$_2$/Si surface.
		}\label{algo}
	\end{figure*}

In this work, we introduce a phase-field modeling framework for Si-based qubit MOSFETs using three different physics modules in COMSOL [44]. This framework employs a full-wave treatment of electrons at cryogenic temperatures, which solves the 2-D Poisson equation, coupled with the tine-independent Schrodinger equation. Using incident, reflected, and transmitted wave functions, we calculate tunneling and exchange coupling. Additionally, we examine the impact of SiO$_2$/Si surface trap distributions on qubit performance. We optimize the plunger gate, barrier gate, gap oxide, gate oxide, and spacer oxide to enhance the tunneling and exchange coupling among quantum dots.

 The following approximations are made in this work.
 
1) We do not consider the magnetic field in this modeling, as this work aims to analyze the electrostatics part of the qubit.
2) We do not include the impact of external noise, coherence, and dephasing. An accurate incorporation of spin effects, noise and coherence requires solving the time-dependent Schrödinger equation with a full Hamiltonian (including magnetic component) converging at cryogenic temperatures, which is beyond the scope of this work.
We consider a 2-D geometry which is also experimentally reported in the recent study by Intel, a Si-bulk MOSFET was used to build a qubit device [5]. A 2-D structure leads to better convergence and reliable results. For the 3-D system, one needs to solve 2-D Schrödinger with 3-D Poisson's including wave treatment which might not lead to a stable simulation platform. Furthermore, the device electrostatic such as the impact of gate field, oxide thickness, and trap impact will follow a similar trend in 3-D as in the 2-D structure.

We analyze the electrostatics part of a small qubit device and engineer the core device for optimal tunnel coupling and electrostatic integrity. This study can help to understand the fundamental ways to optimize and design an efficient qubit for a strong tunnel and exchange coupling.

\section{Methodology}
Fig. \ref{fig:schm} depicts the schematic of a Si-based qubit MOSFET featuring two quantum dots (QD). Our earlier work used the COMSOL Multiphysics module can be used to model complex partial differential equations in semiconductor physics [45], [46]. Fig. \ref{algo} illustrates the phase-field modeling algorithm utilized in this study. We integrated three different physics modules within COMSOL using the Multiphysics tool.

\subsection{Coefficient form Partial Differential Equation Module: Poisson's Equation}
The 2-D Poisson's equation was implemented in the Coefficient from the Partial Differential Equation (PDE) module of the COMSOL. The electron charge density was calculated from the wave function and Fermi-Dirac distribution (Eq. 3). The wave function ($\psi$) and potential function ($\phi$) are defined as dependent variables that are fed back into the Schrödinger module.
\subsection{Schrödinger Module-1: SCHRP Solver}
The COMSOL has an inbuilt Schrödinger module which can be used to calculate the eigen energy and wave function. There are two key steps for the convergence at cryogenic temperatures:

\begin{figure}[!t]
		\centering \hspace{-3mm}
		\includegraphics[width=0.5\textwidth]{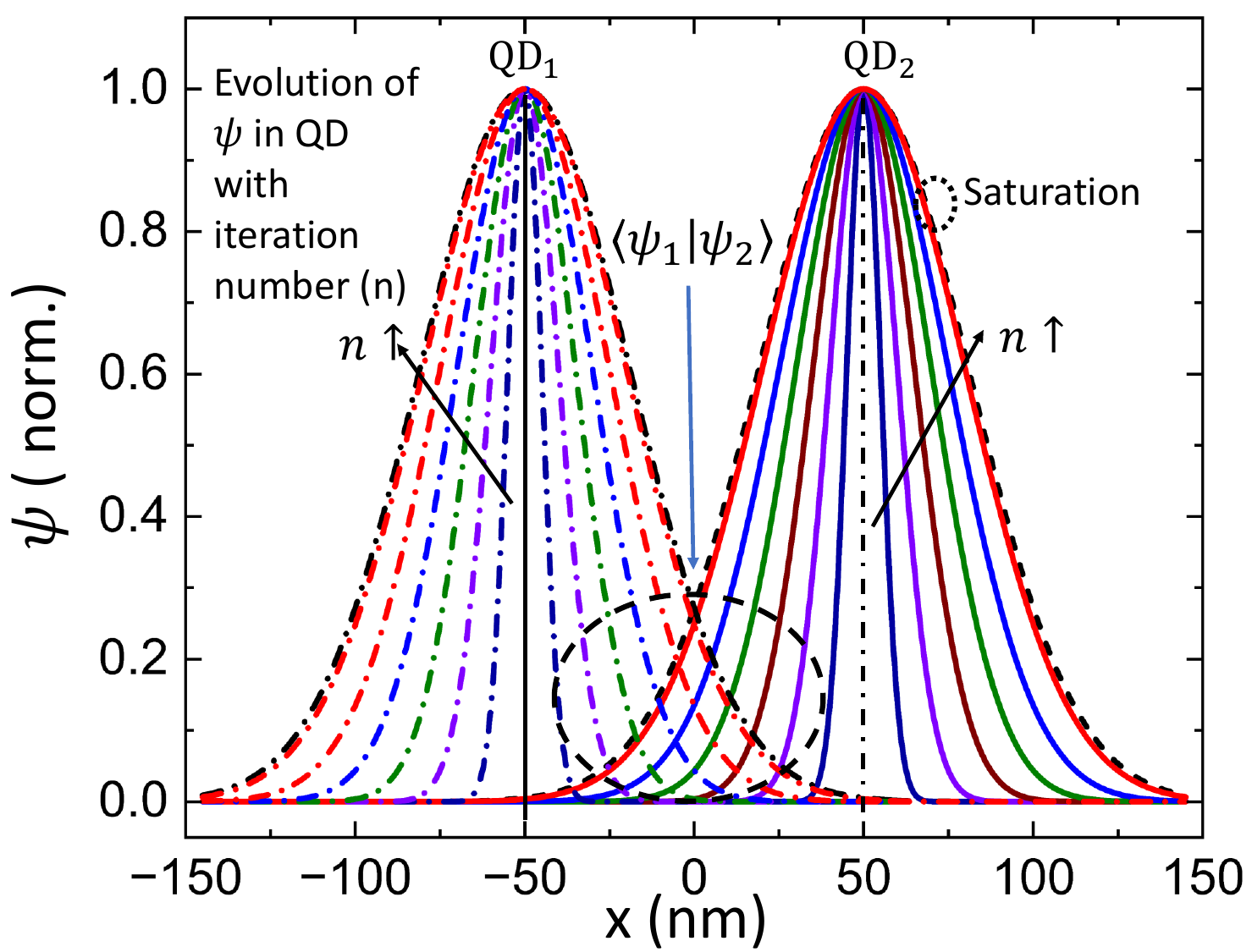}
		\caption{Wave function convergence with the number of iterations in the COMSOL, Greater than 60 iterations are needed for the saturation (or convergence).}
		\label{fig:psi_norm} 
	\end{figure}
\begin{enumerate}
  \item Initially a Gaussian wave function is assumed which is given as an input in the initial value of the Schrödinger Module. This initial function is defined in the analytical function section and then used in the initial value section of the Schrödinger Module. Therefore, this approximation helps the tool to begin with the initial data points to move towards convergence.
  \item First we obtain the solution at $T \geq$ 200 K in a separate stationary study, which is fed back into the Values of Dependent Variable section of the cryogenic study ($T\leq$ 10 K). This step provides the initial potential function,  charge densities, and wave function to start the cryogenic analysis. 
  \item Subsequently, the dependent variables potential $-q\phi$ and $\psi$ are plugged into the electron potential energy section and dependent variable section of the PDE module, respectively. An iterative procedure is used till convergence is achieved.
\end{enumerate}

The iterative solution of Schrödinger and Poisson's solver (SCHRP) evolves the wave function ($\psi$) as shown in Fig. \ref{fig:psi_norm}. Approximately after 60-65 iterations wave function begins to converge. One might need to change the maximum number of iteration limits in the study section to obtain the correct convergence.

\begin{figure}[!t]
		\centering \hspace{-3mm}
		\includegraphics[width=0.5\textwidth]{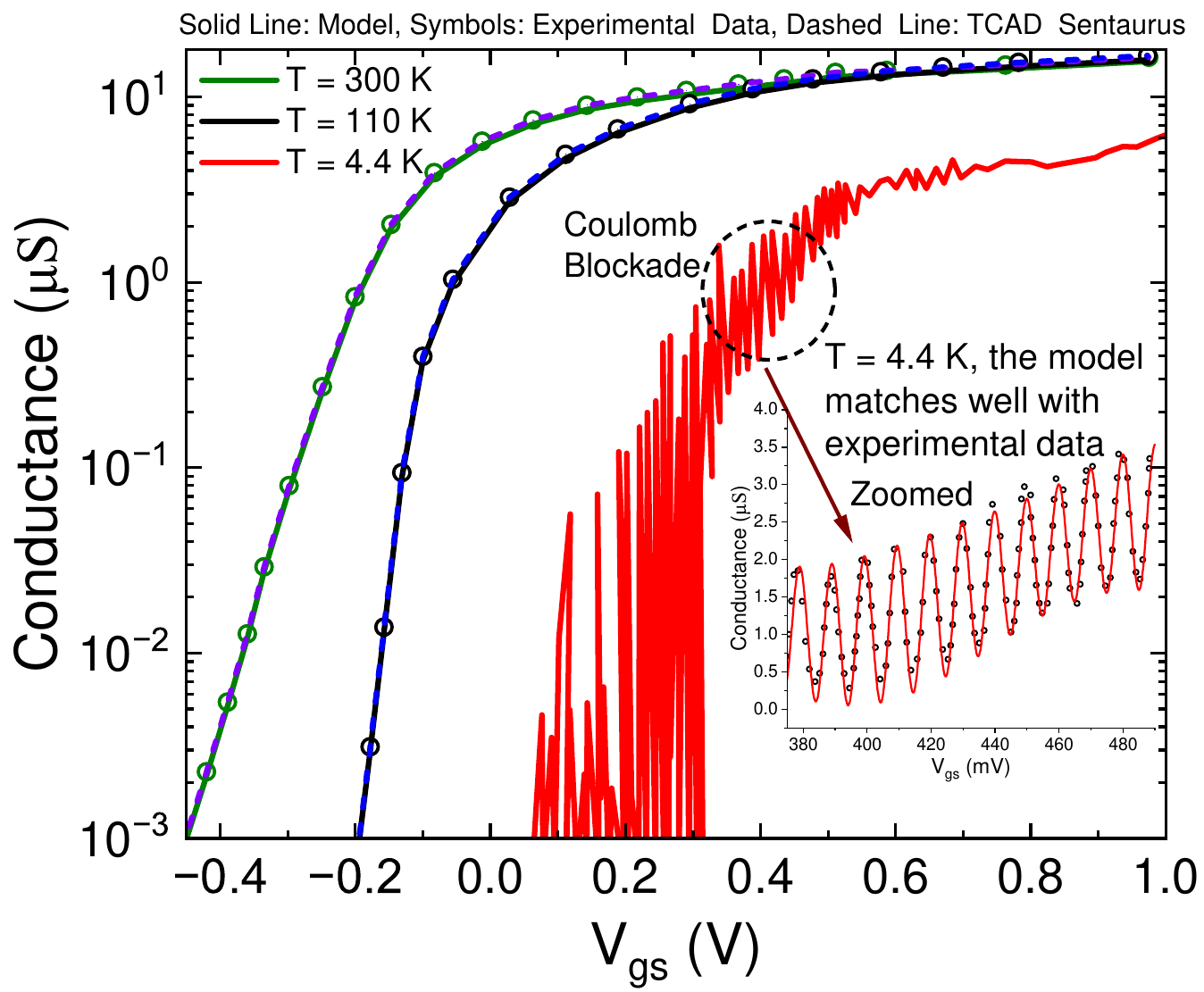}
		\caption{Model validation with experimental data [47] and TCAD Sentaurus software [48]. The model accurately captures experimental Coulomb Blockade oscillations, the essence of qubits. The oscillations are induced due to the Coulomb Blockade with period $\sim \Delta V_{gs}/C_g$. Here, $C_g$ is the gate capacitance.
		}\label{fig:valid} 
	\end{figure}
 
\subsection{Schrödinger Module-2: Full Wave Treatment}
Once we obtain convergence in potential and wave function from the last step, it can be stated that for the given terminal voltages and device physical parameters an accurate potential distribution is obtained at $T \leq$ 10 K. 
The obtained $-q\phi$ is then used in the Electron Potential Energy section of the second 1-D Schrödinger Module to calculate the transmission and reflection coefficient following the exact wave treatment. This $-q\phi$ acts as the electron potential (or barrier) in determining the transmission/reflection coefficients.
COMSOL computes the wave number following Eq. (5), and spatial variation in the wave function with respect to space dimension is calculated from Eq. (6). In the last step the transmission coefficient at each energy step is evaluated by analyzing the transmitted and incident wave magnitude. For simplicity, the tool assumes an initial incident wave magnitude = 1, which evolves with iterations [44].

\subsection{Dependent Variable Transfer between Various Modules}
As explained previously in the above steps the dependent variable transfer is an important step to accurately couple all the modules which can be summarized as:
\begin{enumerate}
  \item PDE: define $\phi$ \& $\psi$ as the dependent variable in the Coefficient form PDE module under the declaration section.
  \item Schrödinger module-1 : Transfer $\phi$ in the electron potential energy section and $\psi$ in the module dependent variable section. Obtain an iterative solution with a proper initial guess (wave function) and solution ($T\geq$ 200 K) [see above steps for detail]. 
  \item The obtained $\phi$ from the above steps is used in the second Schrödinger module to calculate transmission and reflection coefficients with full wave treatment.
\end{enumerate}

 \begin{figure}[!t]
		\centering \hspace{-3mm}
		\includegraphics[width=0.4\textwidth]{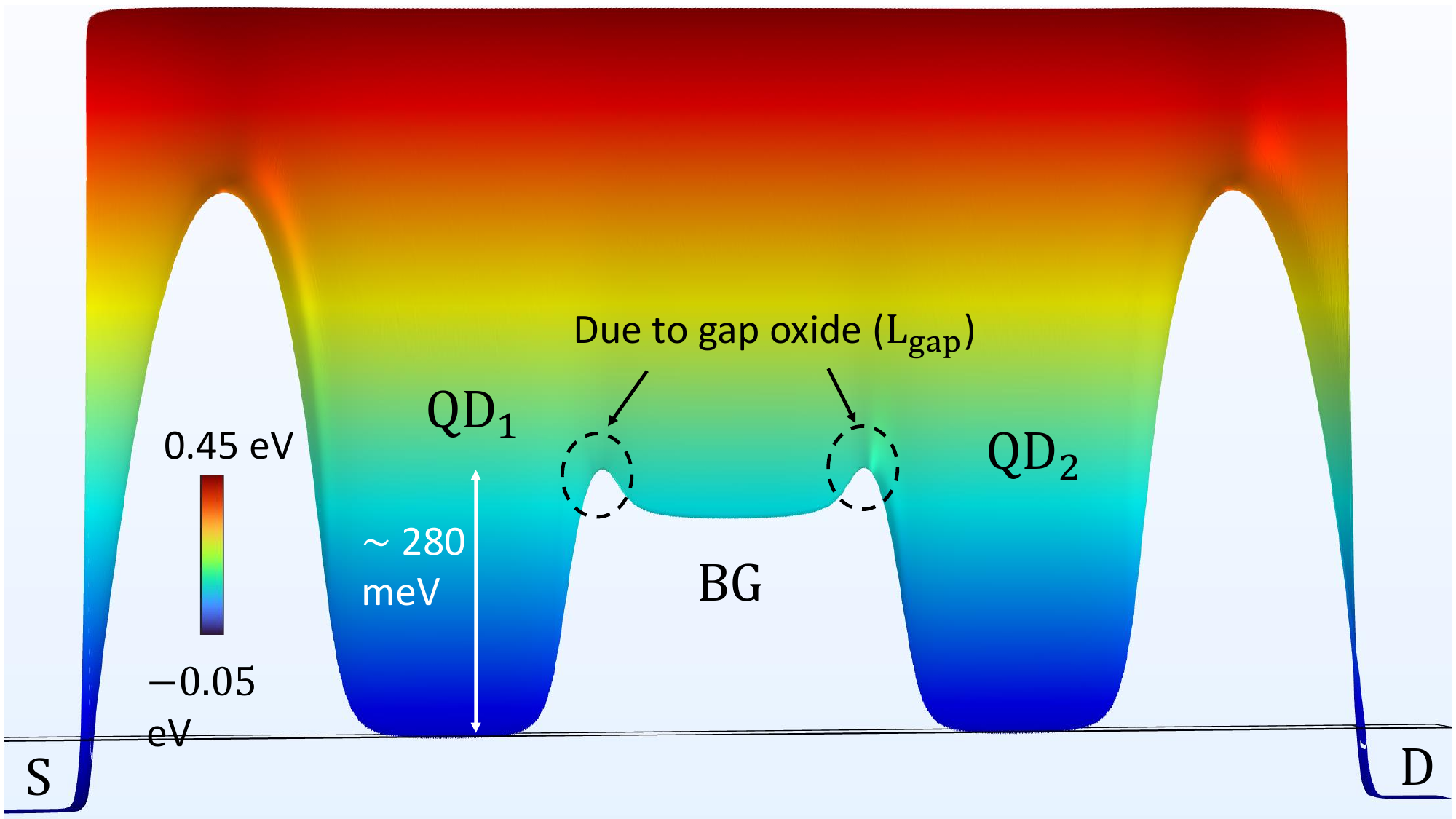}
		\caption{The 2-D surface plot of the conduction band energy with negligible interface trap states shows a QD depth of approximately 280 meV. The local humps near the well boundary result from fringing effects caused by the gap oxides. Parameters from Fig. \ref{fig:schm}}
		\label{fig:Ec_surf} 
	\end{figure}
 \begin{figure}[!t]
		\centering \hspace{-3mm}
		\includegraphics[width=0.4\textwidth]{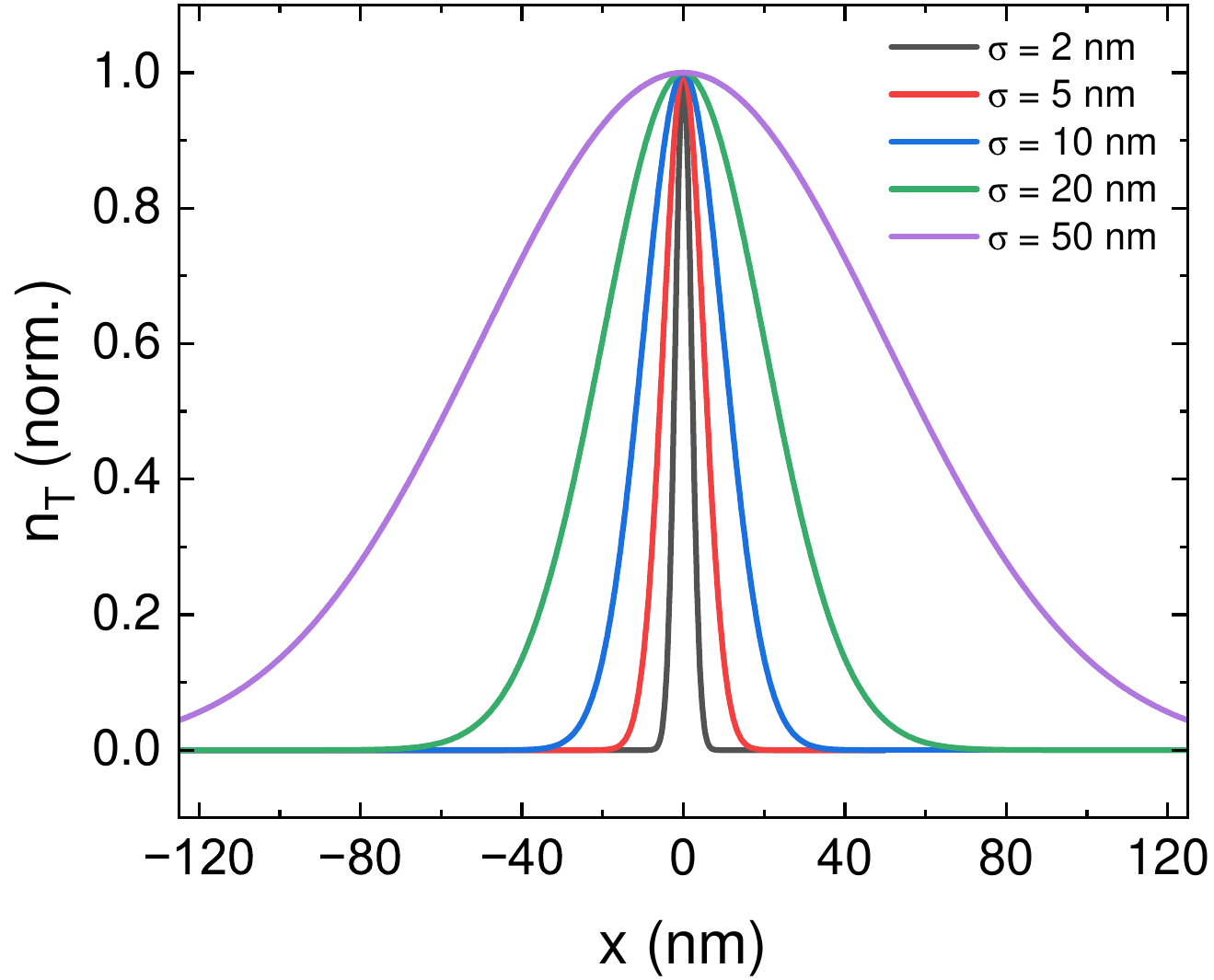}
		\caption{Normalized interface (Si/SiO$_2$) trap distribution centered at the mid-channel for different variance values.}
		\label{fig:nT_sigma} 
	\end{figure}
\section{MODEL VALIDATION}
Fig. \ref{fig:valid} shows the model validations with experimental data reported in [47]. The oscillations in the model have a regular period given by $\Delta V_{gs}=q/\Delta C_{g}$, where $C_g$ is the gate capacitance. The model correctly predicts the temperature dependence, with more pronounced oscillations at lower temperatures due to the reduction in thermal energy overcoming the Coulomb energy barriers, demonstrating its capability to capture single-electron effects characteristic of Coulomb blockade. Furthermore, we also validate the model with standard TCAD Sentaurus [48], proving the additional robustness of the developed phase-field model. 

Note the data in [47] is reported for nano-wire, but the cross-section is a 2-D structure, as shown in Fig. \ref{fig:schm}, leading to matching with the simulations for $T>\geq 110$ K. At cryogenic temperature (4.4 K), a mismatch is observed. However, the simulations still follow the trends capturing oscillations in the $Id-Vg$ which is enough to justify the model validity.
For validation purposes we use the same parameters, as given in [47]. 

 \begin{figure}[!t]
		\centering \hspace{-3mm}
		\includegraphics[width=0.5\textwidth]{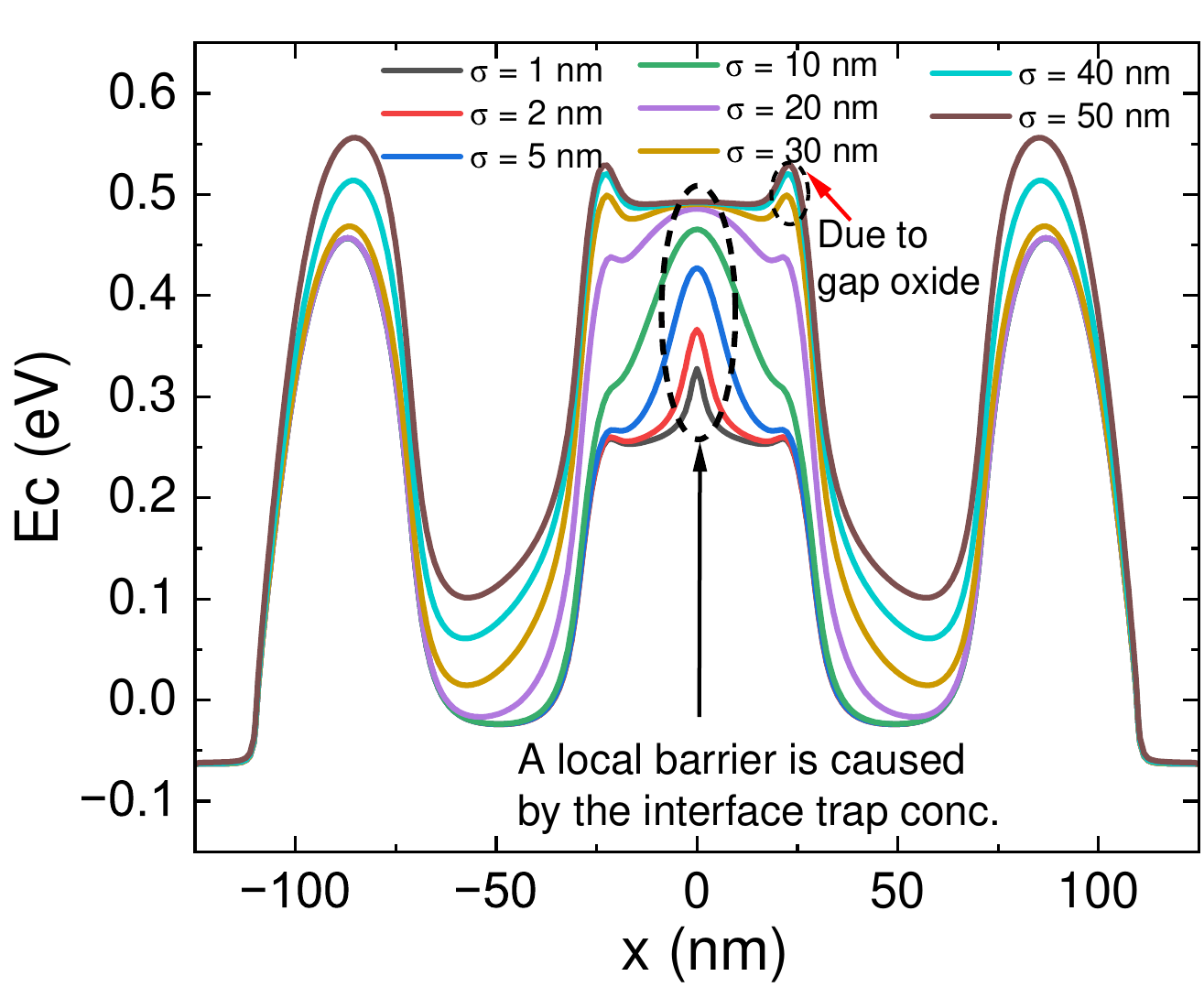}
		\caption{ The conduction band energy is plotted at the Si/SiO$_2$ interface. As $\sigma$ increases, the distribution of interface traps widens, leading to increased local barrier width and height. For $\sigma >$ 30 nm, the maximum barrier height saturates. Typical trap peak concentration is $\sim$ 8$\times$10$^{10}$cm$^{\mathrm{-2}}$, typical values for good interface [49]. Parameters from Fig. \ref{fig:schm}.}
		\label{fig:Ec_sigma} 
	\end{figure}

\section{INTERFACE TRAP and QUBIT PERFORMANCE}
Fig. \ref{fig:Ec_surf} shows the 2-D surface plot of conduction band energy for negligible trap density. The QD's well depth is $\sim$ 280 meV. The local humps near the well boundary are due to the fringing from gap-oxides (L$_{gap}$).
A neutral point defect (trap) is included by assuming the trap’s spatial distribution as a Gaussian with peak concentration $\sim$ 8$\times$10$^{10}$cm$^{-2}$, generally observed in a good interface [49]. Assuming a Gaussian trap distribution is widely used to simulate/model the impact of interface trap density [50], [51], 
Fig. \ref{fig:nT_sigma} shows the typical normalized interface trap distribution centered at the mid-channel for the various variance values. 

\begin{figure}[!t]
		\centering \hspace{-3mm}	\includegraphics[width=0.45\textwidth]{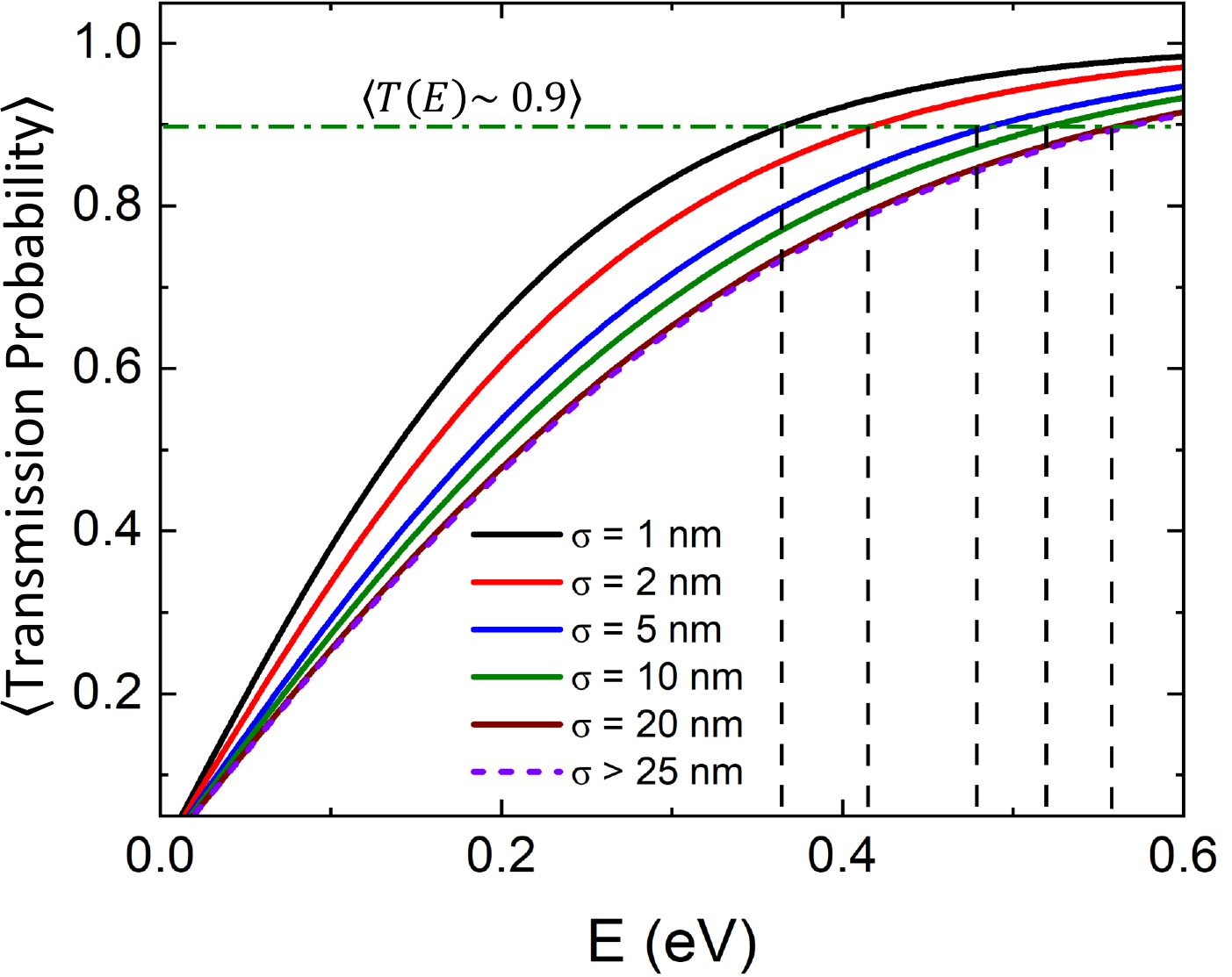}
		\caption{Broader trap distributions increase the tunnel barrier height and width, decreasing tunneling and associated transmission probabilities. Higher gate voltages are required to reduce the tunnel barriers and regain strong quantum mechanical coupling between quantum dots.
		}\label{fig:T(E)_sigma} 
	\end{figure}

 \begin{figure}[!t]
		\centering \hspace{-3mm}	\includegraphics[width=0.4\textwidth]{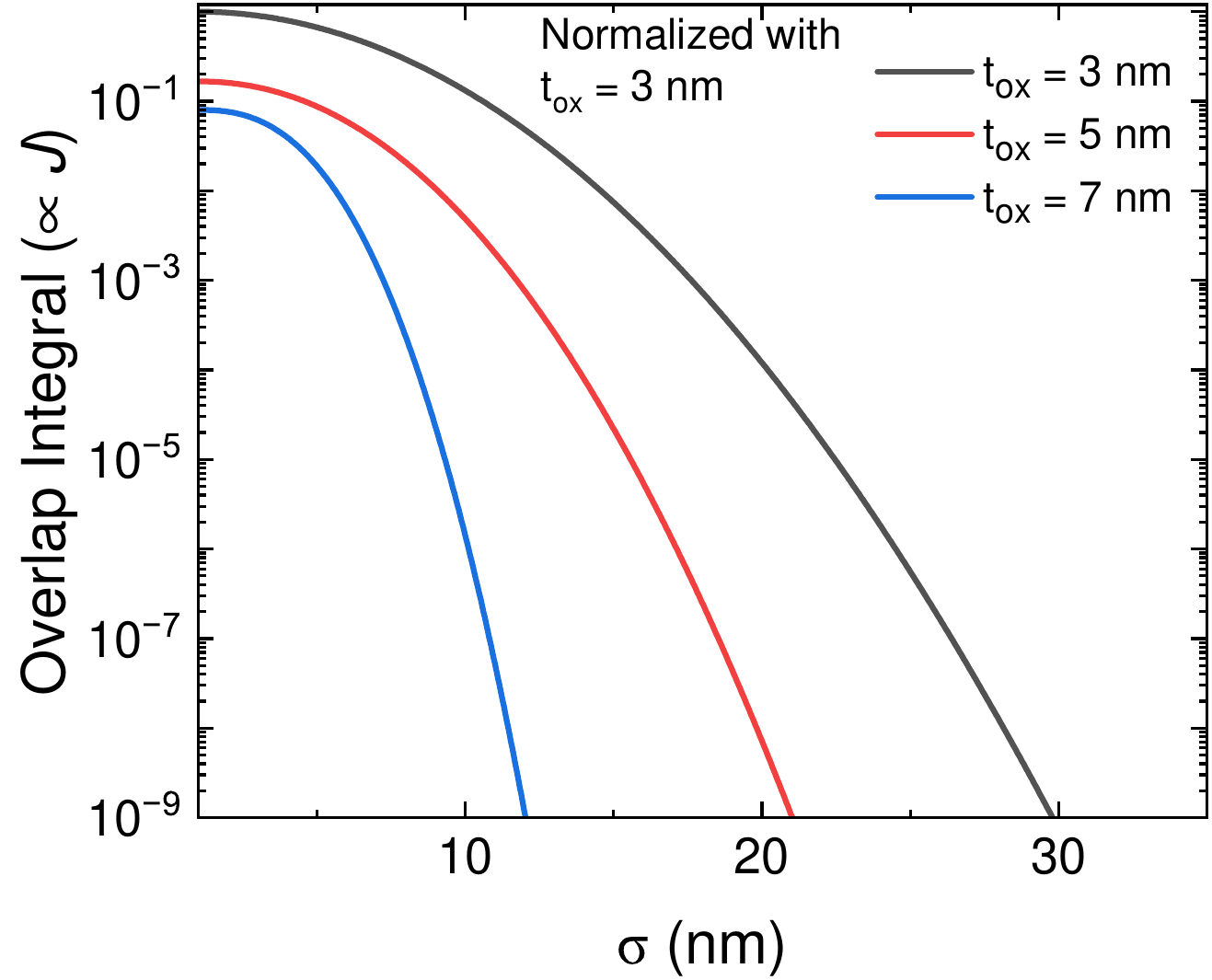}
		\caption{Increasing barrier height reduces wave function overlap $\left (  \left< \psi_1 | \psi_2 \right>\downarrow  \right )$, thereby lowering exchange coupling in quantum dots. Thicker gate oxides further weaken plunger and barrier gate control, reducing exchange coupling.
		}\label{fig:J_sigma_tox} 
	\end{figure}
Since the interface traps are generally electron traps,[49] these traps possess a negative charge when occupied, leading to a local increase in the barrier height of the quantum dot. Fig. \ref{fig:Ec_sigma} plotted at Si/SiO$_2$ interface considers a Gaussian distribution of traps of the fixed peak but standard deviation $\left ( \sigma \right )$ varying from 1 nm to 50 nm. An increasing $\sigma$ increases the total trap charge density, i.e., $\sim N_t\times A_0$. Here $A_0$ is the trap distribution area which increases with $\sigma$, leading to an enhancement in the total trap charge and increasing barrier height, until the trap distribution covers the whole channel region, causing the saturation for $\sigma \geq$ 40 nm. 

  \begin{figure}[!t]
		\centering \hspace{-3mm}	\includegraphics[width=0.45\textwidth]{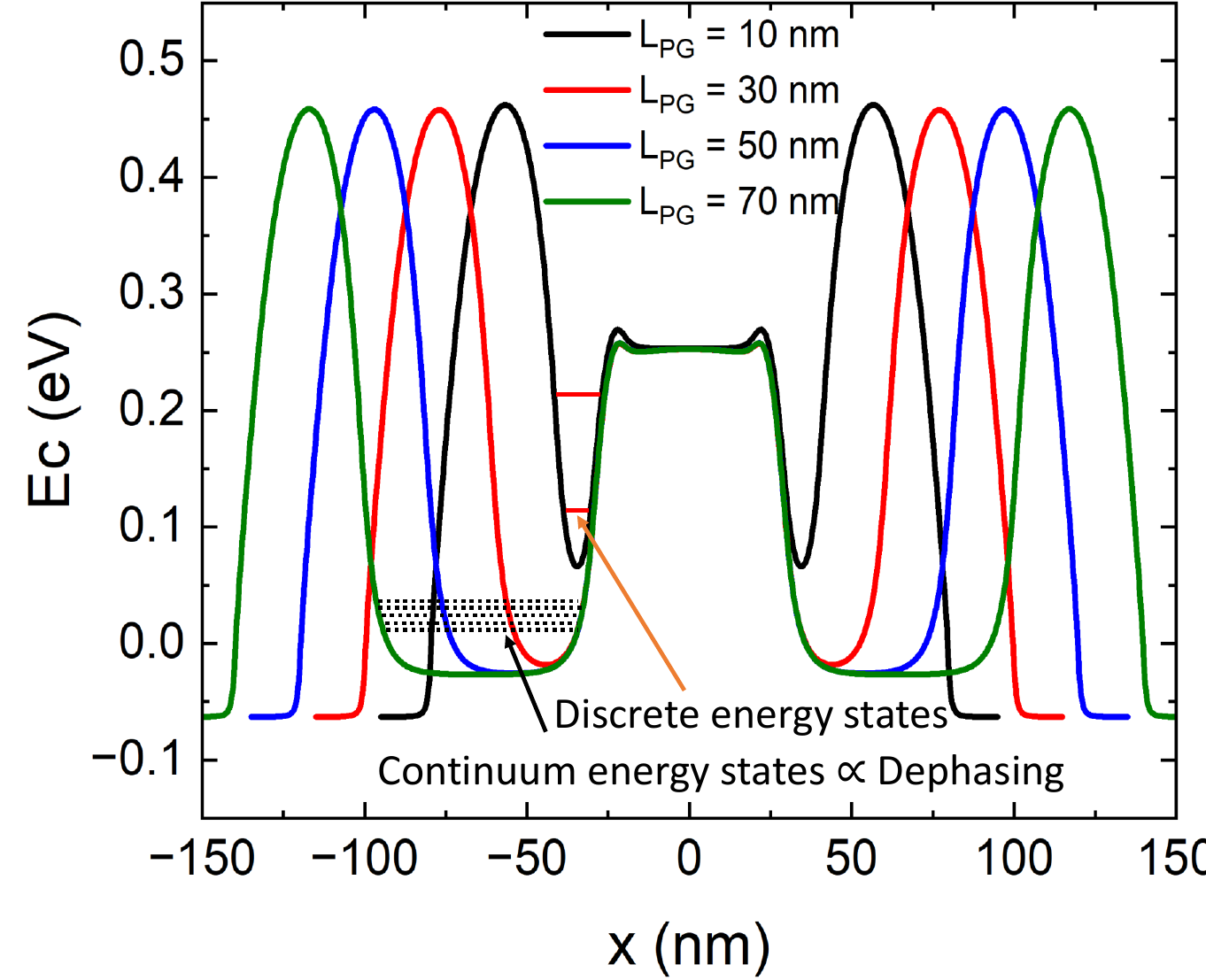}
		\caption{At the larger value of L$_{PG}$, a continuum of energy states is observed, inducing many possible interaction states among QD electron wave functions.
		}\label{fig:Ec_LPG} 
	\end{figure}

\begin{figure}[!t]
		\centering \hspace{-3mm}	\includegraphics[width=0.4\textwidth]{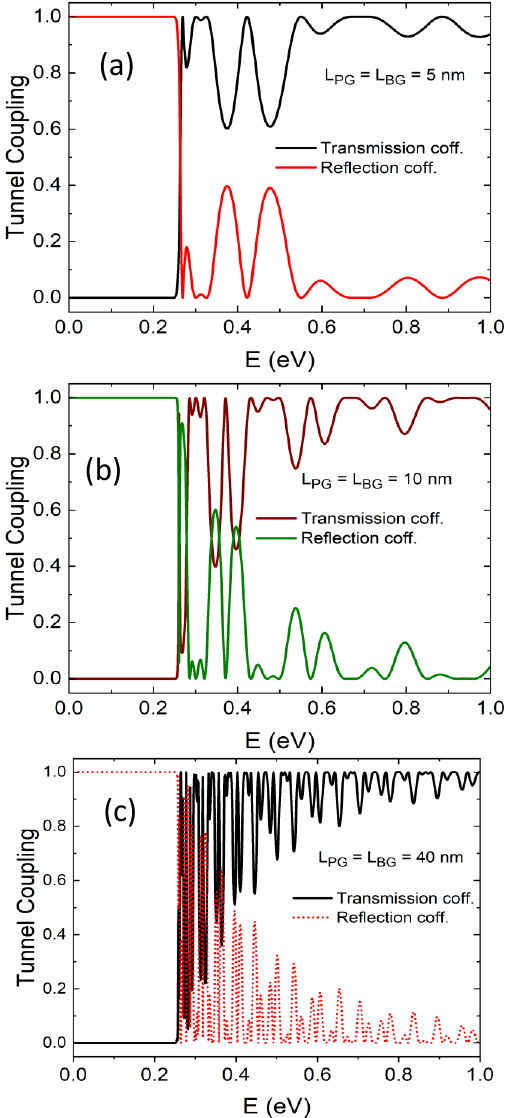}
		\caption{The transmission probability for the plunger (barrier: L$_{BG}$) gate length set to 5 nm, revealing only 2 energy modes due to a high energy separation $ \left ( \Delta E_i \right )$. This results in minimal oscillations in the overlap integral and T(E). As L$_{PG}$ (L$_{BG}$ increases, reducing $\Delta E_i$, a continuum of energy states emerges, markedly increasing oscillations in T(E) and the overlap integral, as observed in (b) and (c).
		}\label{fig:TE_LPG}
	\end{figure}
 \begin{figure}[!t]
		\centering \hspace{-3mm}	\includegraphics[width=0.45\textwidth]{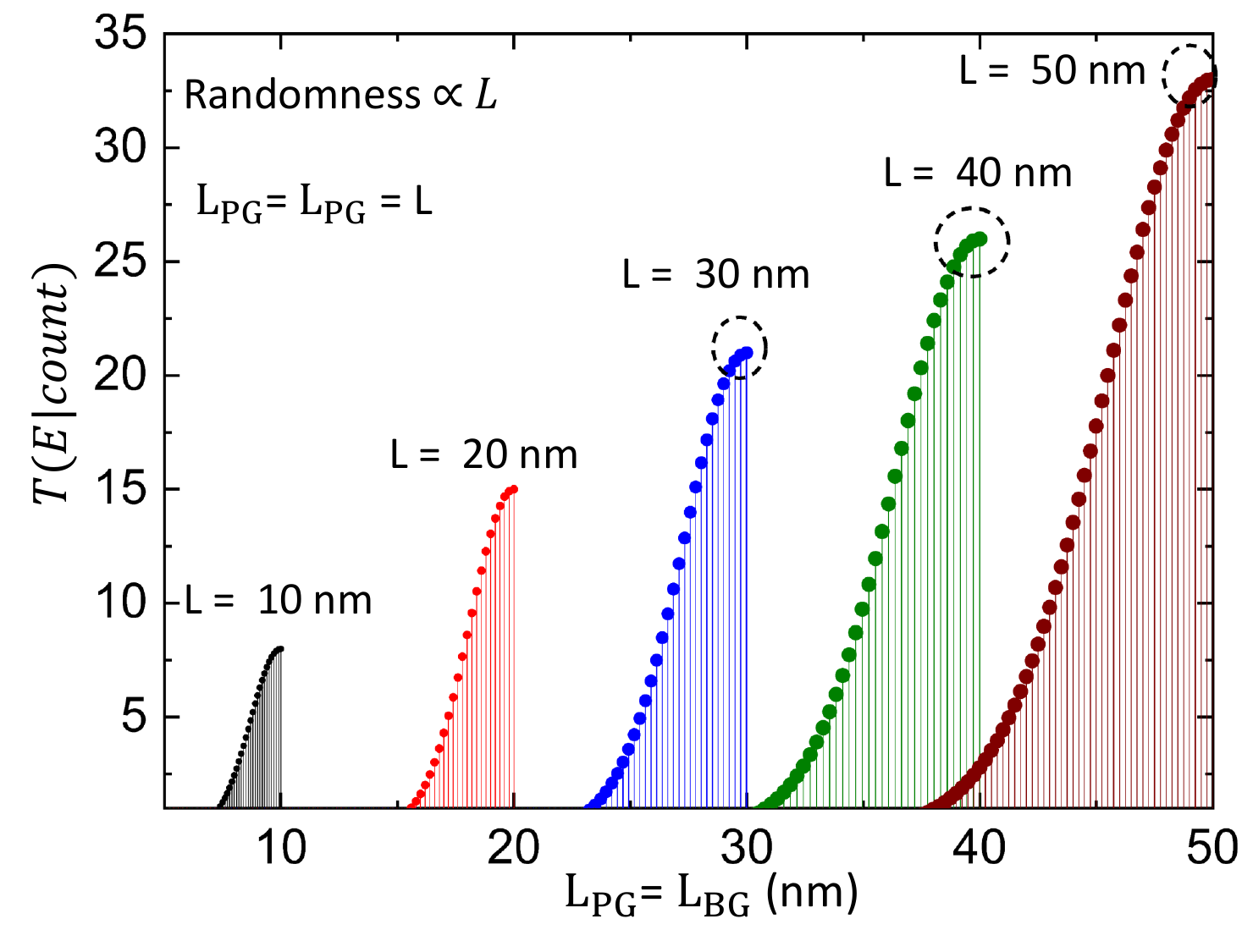}
		\caption{The continuum energy states at large L$_{PG}$ (L$_{BG}$) increase the probability of unintended electron transitions between states, introducing randomness in the tunnel coupling. The variance of this randomness also increases with increasing L$_{PG}$.
		}\label{fig:J_random} 
	\end{figure}
 \begin{figure}[!t]
		\centering \hspace{-3mm}	\includegraphics[width=0.45\textwidth]{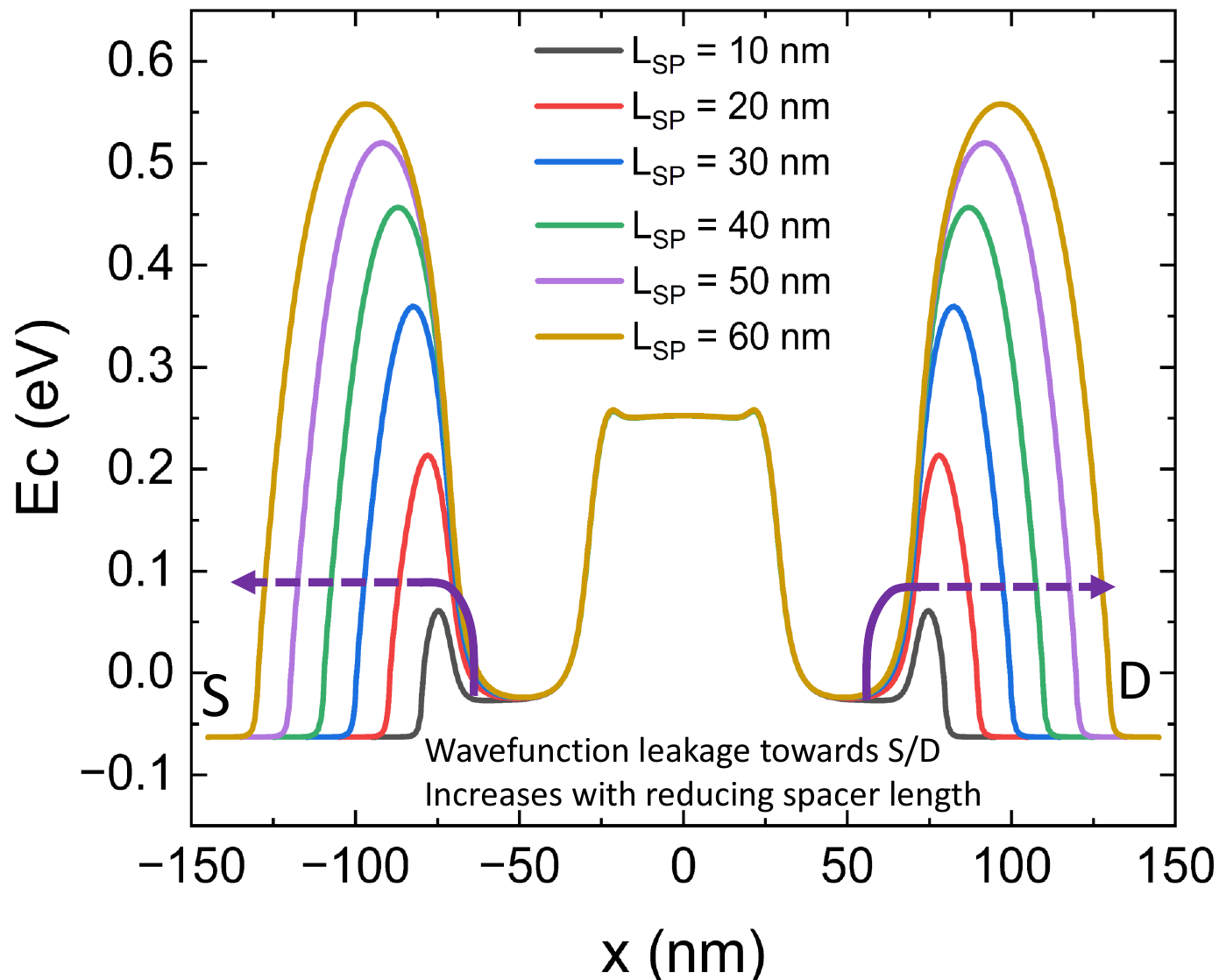}
		\caption{For L$_{SP}<$  30 nm, source/drain depletion alters well depths: L$_{SP}<$10 nm causes quantum well disappearance due to source/drain degeneracy, shifting control to depletion width. Here $L_{SP}$ is the S/D spacer length.
		}\label{fig:Ec_LSP} 
	\end{figure}

Fig. \ref{fig:T(E)_sigma} shows the impact of trap distribution on the average tunnel coupling of the quantum dots. The increasing barrier height caused by the higher $\sigma$ reduces the tunnel coupling. Furthermore, the energy maxima at which T(E) $\sim$ 0.9 increases with barrier height. Therefore, the energy required (gate voltage) to strongly couple such quantum dots will increase with the broader trap distribution. 
\begin{figure}[!t]
		\centering \hspace{-3mm}	\includegraphics[width=0.45\textwidth]{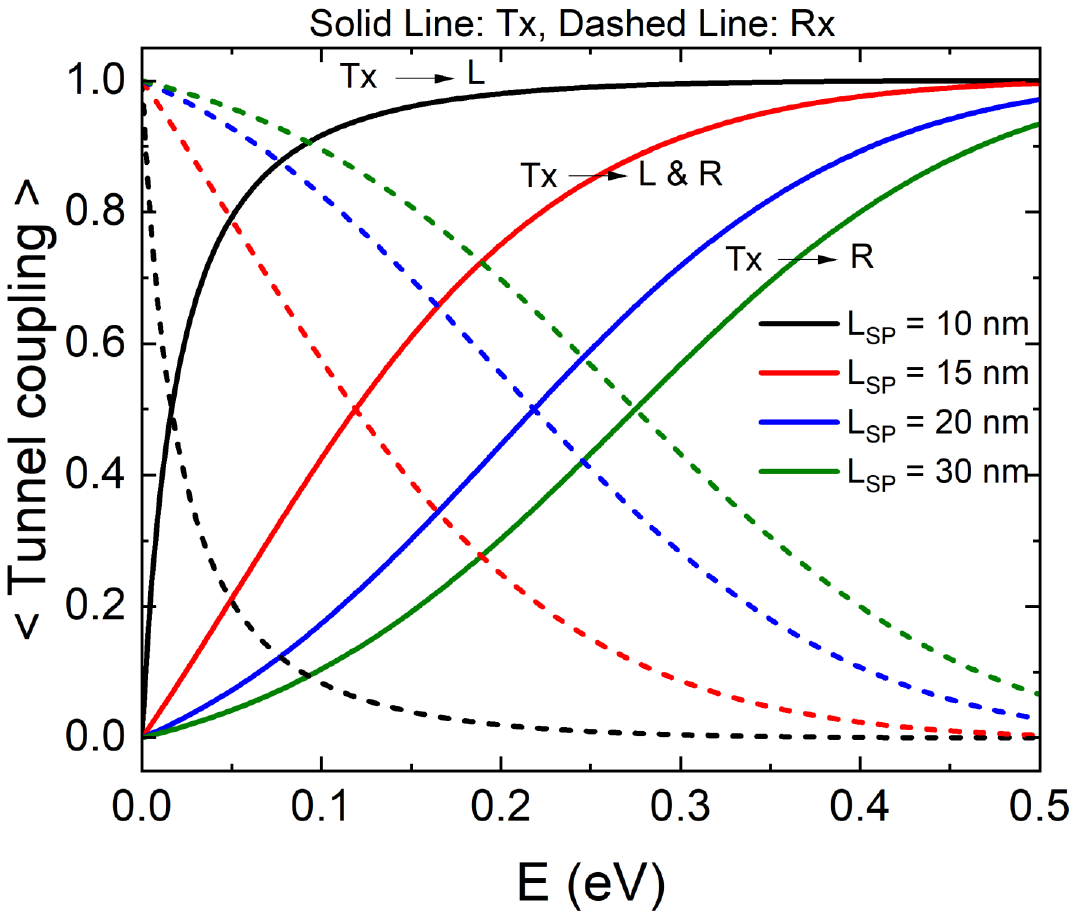}
		\caption{At LSP = 10 nm, most electrons flow to the source and drain. At L$_{SP}$ = 15 nm, 50\% of wave function diffuses to S/D. At L$_{SP}>$  30 nm there is only $\sim$8 \% leakage to S/D, emphasizing the need for smaller spacer widths for effective quantum dots.
		}\label{fig:Tx_LSP} 
	\end{figure}
 \begin{figure}[!t]
		\centering \hspace{-3mm}	\includegraphics[width=0.45\textwidth]{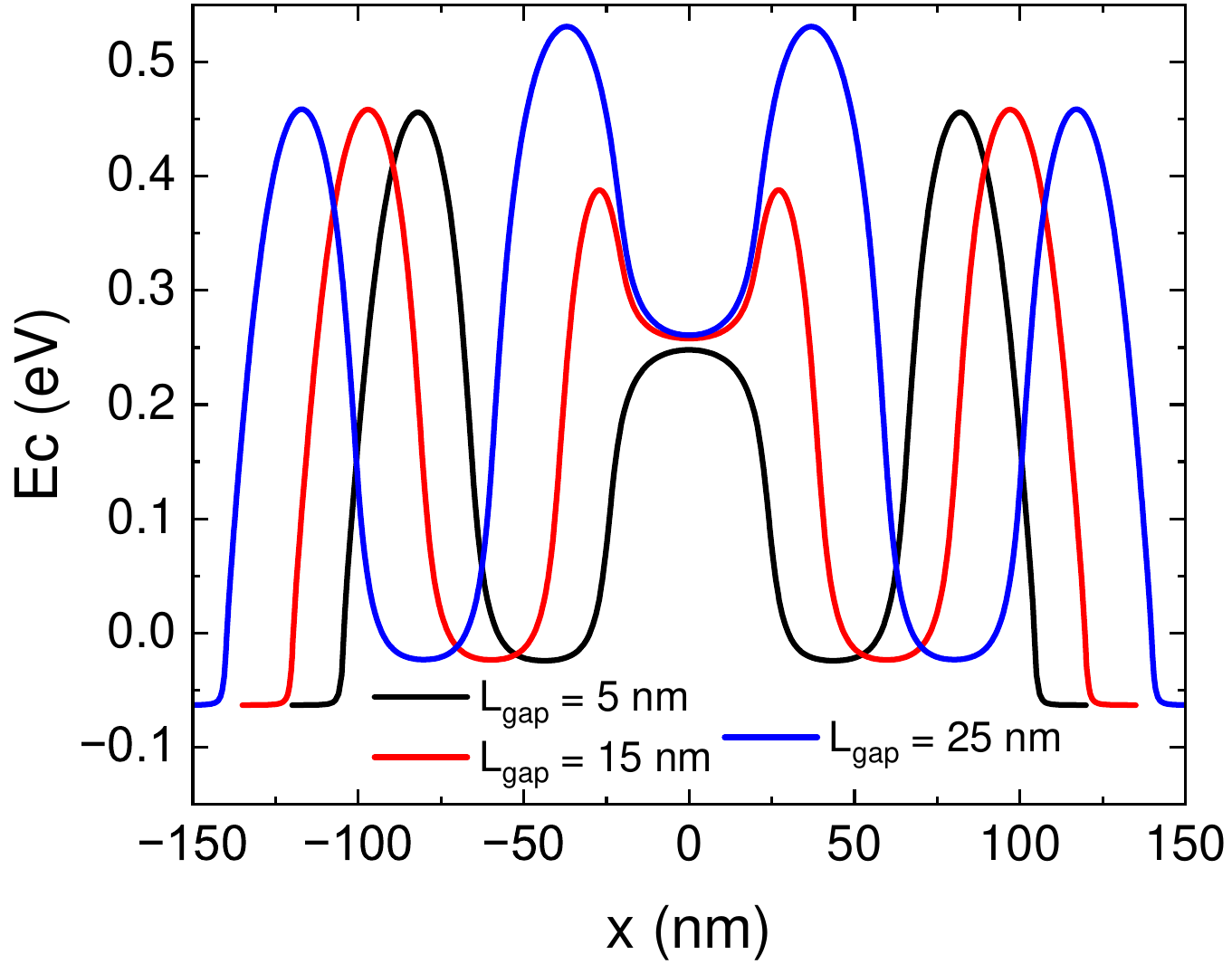}
		\caption{Increasing L$_{gap}$ induces corner humps in the barrier from local fringing fields, raising the barrier height of the quantum wells. At very large L$_{gap}$ (25 nm), wells might not interact due to the significant potential barrier.
		}\label{fig:Ec_Lgap} 
	\end{figure}
The impact of trap distribution on exchange coupling is shown in Fig. \ref{fig:J_sigma_tox}. Increasing barrier height reduces the wave function’s overlap integral $< \psi_1| \psi_2>$, leading to a reduction in the exchange coupling. Furthermore, the thicker the gate oxide, the weaker the plunger and barrier gate control, further reducing the exchange coupling. 

 \begin{figure}[!t]
		\centering \hspace{-3mm}	
  \includegraphics[width=0.45\textwidth]{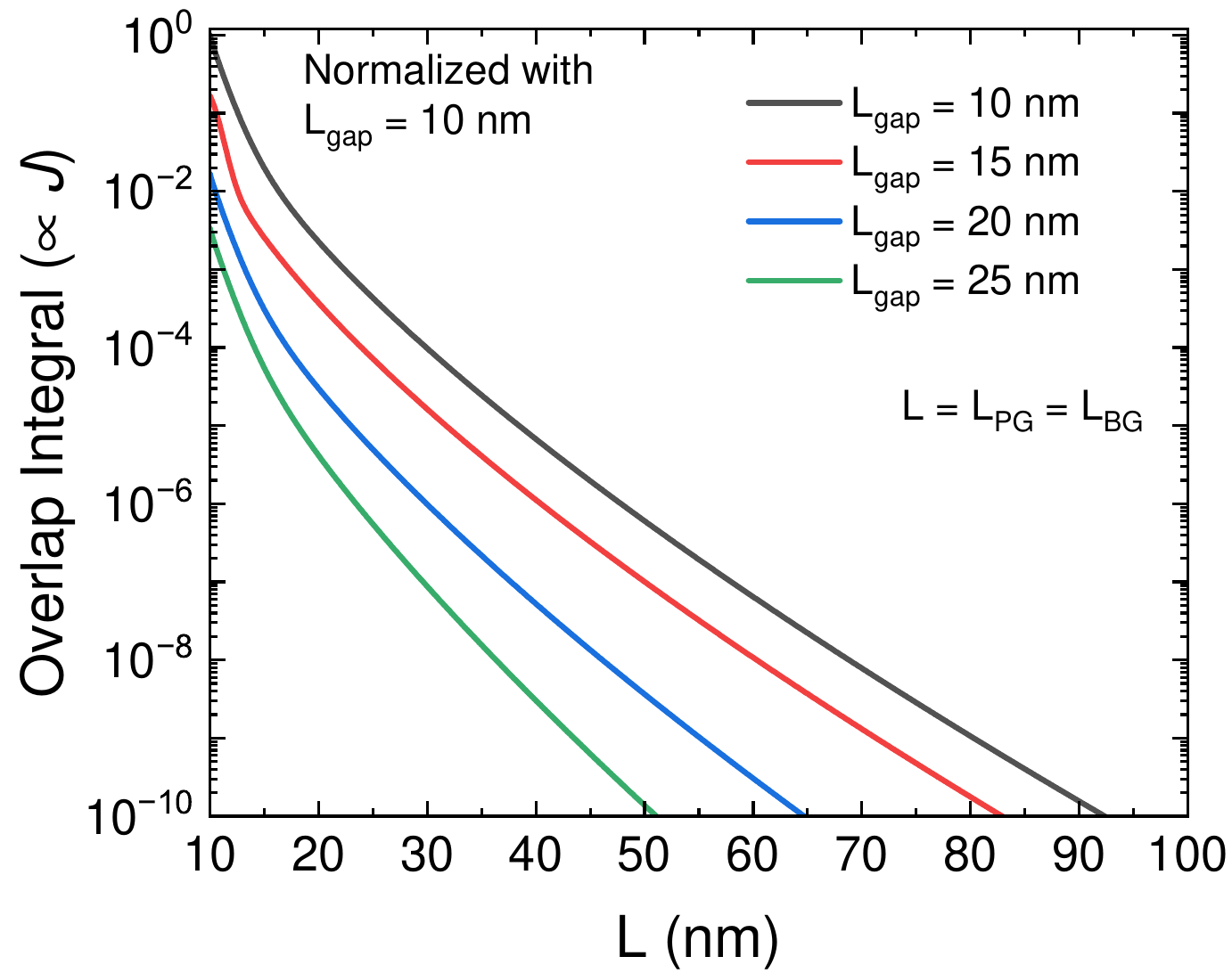}
		\caption{Exchange coupling decreases with increasing gap length due to the increased barrier height and width and resulting decreased wave function overlap.
		}\label{fig:J_Lgap} 
	\end{figure}
 \begin{figure}[!t]
		\centering \hspace{-3mm}	\includegraphics[width=0.5\textwidth]{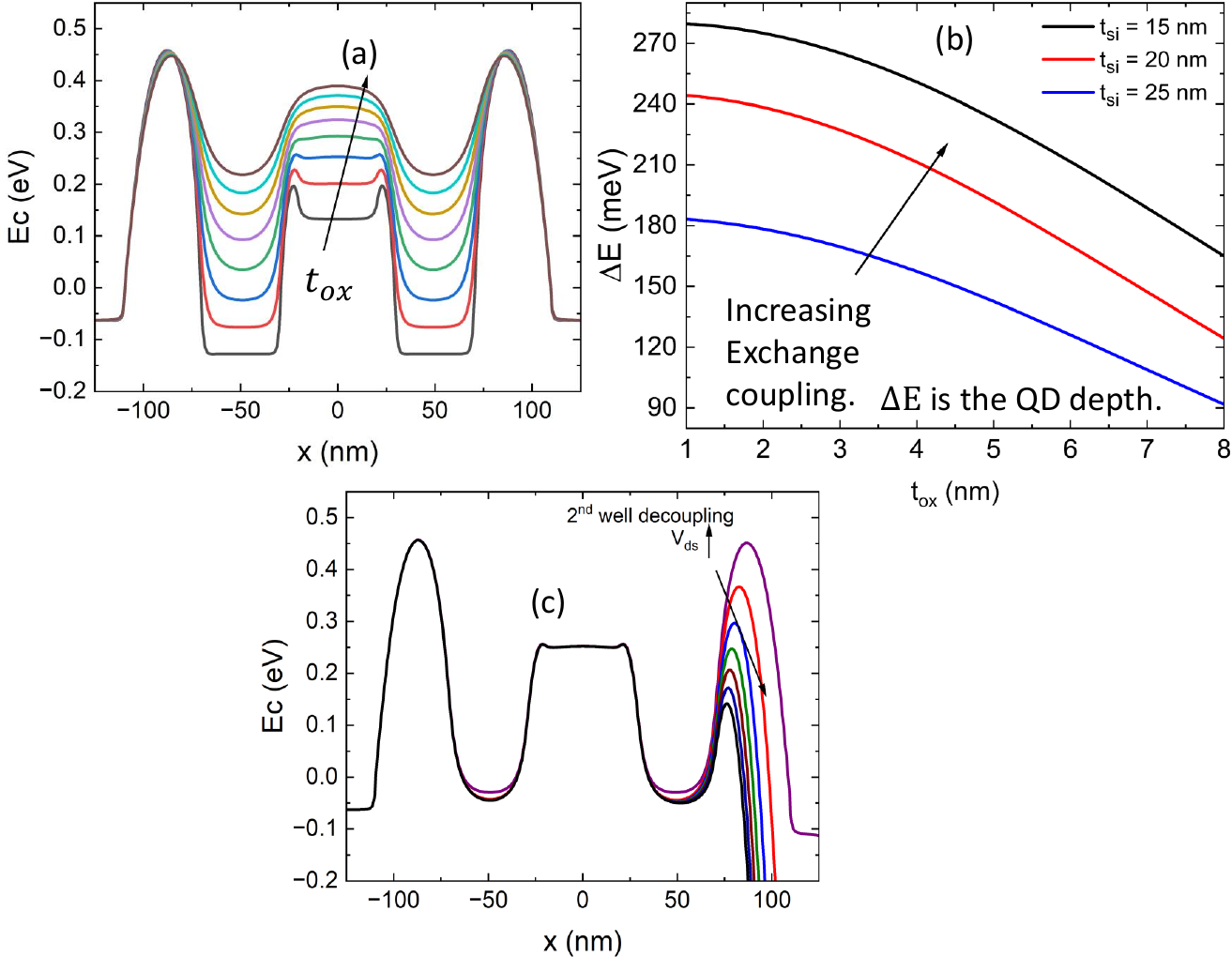}
		\caption{(a) Increasing gate oxide thickness reduces the well depth and gate control, diminishing tunnel coupling. (b) Increasing silicon film thickness also reduces well depths due to diminished gate control. This reduction in the quantum well depth increases the unwanted wave function leakage, decreasing tunnel coupling. (c) The drain bias affects quantum well interactions by modulating the well near the drain side. Increasing drain voltage causes drain-induced barrier lowering, which reduces the well depth near the drain side, reducing exchange coupling.  
		}\label{fig:Ec_tox_tsi} 
	\end{figure}
\section{PLUNGER and BARRIER GATE ENGINEERING}
Fig. \ref{fig:Ec_LPG} shows the impact of plunger gate length (L$_{PG}$) on the quantum well. The quantum well depth shows a negligible dependency for L$_{PG}>$ 30 nm. However, the well depths decrease at the smaller plunger gate lengths. At a very small value of L$_{PG}<$ 10 nm, the eigen-energy separation between dot electrons is exceptionally high $ \left ( \Delta E_i\propto 1/L^2 \right )$. Therefore, there are only a few available states at which quantum dots can interact with each other.

Fig. \ref{fig:TE_LPG}(a) shows the transmission probability for plunger (L$_{PG}$) and barrier (L$_{BG}$) gate length = 5 nm. Due to the very high $\Delta E_i$  only two modes (energy states) are available, leading to small oscillations in the overlap integral and T(E). The increasing value of L$_{PG}$ (L$_{BG}$) reduces the $\Delta E_i$ causing a continuum of energy states, which significantly increases the oscillations in the T(E), as shown in (b) \& (c). This increase in oscillations increases the likelihood of unintended electron transitions between states, leading to randomness in the tunnel coupling, as shown in Fig. \ref{fig:J_random}. The randomness is calculated as the number of oscillations in the transmission coefficient till it saturates with respect to energy.
The variance of randomness also increases with L$_{PG}$.  

\section{SPACER and GAP OXIDE ENGINEERING}
Fig. \ref{fig:Ec_LSP} shows the conduction band profile of the quantum wells for various source/drain spacer lengths (L$_{SP}$). For the L$_{SP}<$ 30 nm, the source/drain depletion region begins to alter the well depths. At very small L$_{SP}<$  10 nm, the quantum well begins to disappear due to the degeneracy of S/D, and the region below the plunger gates is now controlled by the S/D depletion width. Its impact on tunnel coupling is shown in Fig. \ref{fig:Tx_LSP}. For L$_{SP}$ = 10 nm, almost the entire wave function of dot electron is transmitted towards the source and drain side from QD${_1}$ and QD${_2}$, respectively. At L$_{SP}$ = 15 nm, almost 50\% of the electron wave function diffuses towards S/D. For L$_{SP}>$  30 nm, there is $\sim$ 8 \% leakage of wave function towards S/D, and almost the entire quantum well is controlled by the plunger/barrier gates. Therefore, keeping a larger S/D drain spacer can help to enhance quantum dot coupling. 

Fig. \ref{fig:Ec_Lgap} shows the interface conduction band energy (QD) for the various gap oxide lengths (L$_{gap}$). Here L$_{gap}$ is the oxide gap length between adjacent plunger/barrier gates (see Fig. \ref{fig:schm}). The increasing value of L$_{gap}$ induces corner humps in the barrier due to the local fringing fields. These local humps increase the barrier height of the quantum wells. At very large L$_{gap}$, it might be possible that wells do not interact with each other due to a very large potential barrier. Fig. \ref{fig:J_Lgap} shows the exchange coupling for the various gap lengths. Increasing L$_{gap}$ reduces exchange coupling due to decreased overlap between the electron wave functions. 

Fig. \ref{fig:Ec_tox_tsi}(a) shows the impact of gate oxide thickness on quantum dots. Increasing gate oxide thickness reduces the gate control and well depths, which reduces the exchange coupling. The increasing value of silicon film thickness also reduces the quantum well depths due to reduced gate control, as shown in Fig. \ref{fig:Ec_tox_tsi}(b). An increasing drain voltage value causes drain-induced barrier lowering, reducing the quantum well depth near the drain side, as shown in Fig. \ref{fig:Ec_tox_tsi}(c), which will decrease the exchange and tunnel coupling.

\begin{figure}[!t]
		\centering \hspace{-3mm}	\includegraphics[width=0.5\textwidth]{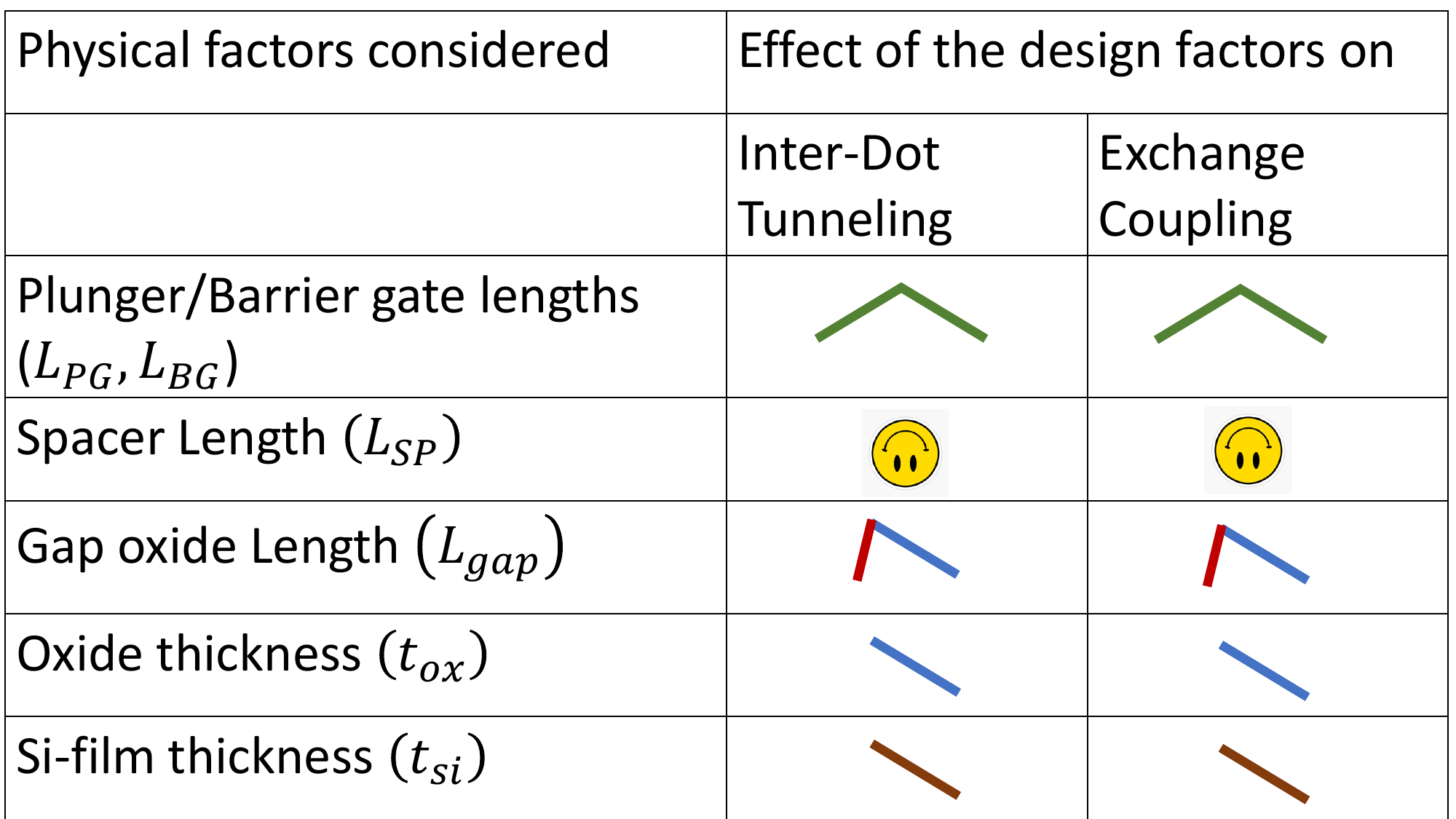}
		\caption{Summary of the outcome of this work. For an intermediate barrier/plunger length coupling is maximum. Too small and too large spacer length reduces the coupling. At very small gap oxide the wells will merge, leading to zero coupling, at very large gap oxide coupling also reduces. Increasing gate oxide and Si film thickness reduces coupling.
		}\label{fig:table_1} 
	\end{figure}

Fig. \ref{fig:table_1} shows the Summary of this work: 
The coupling is maximized at an intermediate barrier/plunger gate length. Too small or too large spacer lengths reduce the coupling. When the oxide gap is very small, the quantum wells merge, resulting in zero coupling, and at very large oxide gaps, coupling also decreases. Increasing the gate oxide thickness and the Si film thickness decreases the coupling.

\section{Conclusion}
 We have elucidated critical factors affecting Si-based qubit MOSFET performance by integrating electrostatics, quantum mechanics, and magnetic effects, including full wave treatment of electrons at cryogenic temperatures. Our findings highlight the significant impact of interface traps on barrier heights which influence tunnel coupling. Optimal engineering of plunger and barrier gates, spacer widths, and gap oxide lengths was shown to enhance quantum well depth control and minimize unwanted wave leakage, increasing the coupling and electrostatic integrity.

\end{document}